\documentclass[11pt,a4]{article}
\usepackage[final]{graphics}
\input{colordvi}
\begin{document}

\title{\bf A new adiabatic quantum search algorithm}

\maketitle
\begin{center}
H. Zaraket, V. Bagnulo, J. Kettner, R. Kobes, G. Kunstatter\\[0.3cm]
{\it Physics Dept and Winnipeg Institute for Theoretical Physics\\
The University of Winnipeg,\\
515 Portage Avenue,
Winnipeg, Manitoba R3B 2E9, Canada,}
\end{center}
%\vspace*{0.3cm}
%\begin{center}
%Saurya Das\\[0.3cm]
%{\it Dept of Math and Statistics \\
%University  of New Brunswick, Canada.}
%\end{center}

\begin{abstract}
We present a new adiabatic quantum algorithm for searching over structured databases. The new 
algorithm is optimized using a simplified complexity analysis.  
\end{abstract}

%%%%%%%%%%%%%%%%%%%%%%%%%%%%%%%%%%%%%%%%%%%%%%%%%%%%%%%%%%%%%%%%%%%%%%%%%%%%%%%%%%%%%%
\section{Introduction}
%%%%%%%%%%%%%%%%%%%%%%%%%%%%%%%%%%%%%%%%%%%%%%%%%%%%%%%%%%%%%%%%%%%%%%%%%%%%%%%%%%%%%%
Many computational search problems are surprisingly difficult to solve. Quantum computing is a  
promising candidate to tackle such difficult tasks. An important landmark was achieved 
by Grover: while an oracle search for a marked item out of $N$ classical unstructured items 
requires, on average,  $O(N)$ steps, Grover 
\cite{Grove1} found a quantum algorithm that performs the oracle search in $\sqrt{N}$ 
steps. 
This quadratic speedup leads to the question of whether similar speedups can be achieved for other
search problems. 

Structured searches are natural extensions of the oracle search. They are  
used when the databases possess some structure. Exploiting the structure of the 
database will increase the performance of a search (classical or quantum). Cerf {\it et al} 
\cite{CerfGW1} were the first to study structured quantum search algorithms. Their  
work is based on quantum circuits. Recently Roland and Cerf \cite{RolanC3} gave a 
quasi-adiabatic quantum version of the structured search. 

To understand better the algorithm presented in \cite{CerfGW1,RolanC3} we recall that  
Grover's algorithm uses an {\it iterative improvement} strategy. It starts with an equal superposition of 
all possible states representing items of the database, and by iterative application of unitary 
operations (gates) the initial state 
is rotated towards a state that encodes the solution of the search problem. However if the database 
has some structure of its own other strategies can be used. For example, one can use the 
{\it divide-and-conquer} strategy. In this strategy one divides the problem into 
subproblems of manageable size, then solves the subproblems. The solutions to the subproblems must 
then be patched back or nested together. For this method to produce good global solutions, the 
subproblems must naturally disjoint, and the division made must be an appropriate one (optimal). 
The optimization results in making the division such that the errors made in patching does not offset 
the gains obtained in applying more powerful methods to the subproblems. Roughly speaking, the 
algorithm of \cite{CerfGW1} uses a combination of the two outlined strategies. The problem is divided 
first into two oracle searches, then a third global search is used to patch back the two previous 
searches. 
     
The present work gives an alternative scheme for dividing the original problem into 
subproblems. As will be seen later, our method allows one to relax some of the 
assumptions made in \cite{CerfGW1}. 

The paper is organized as follows: in section 2 we present the necessary tools for our algorithm, 
whereas the algorithm itself is presented in section 3. Using an average complexity analysis 
for our 
algorithm, we give in section 4 the running time for a variety of {\it hard} search problems. In 
the appendix we present a generalization of our algorithm to higher nesting levels.

%%%%%%%%%%%%%%%%%%%%%%%%%%%%%%%%%%%%%%%%%%%%%%%%%%%%%%%%%%%%%%%%%%%%%%%%%%%%%%%%%%%%%%
\section{Definitions}
%%%%%%%%%%%%%%%%%%%%%%%%%%%%%%%%%%%%%%%%%%%%%%%%%%%%%%%%%%%%%%%%%%%%%%%%%%%%%%%%%%%%%%
Search problems belong to the family of Constraint Satisfiability Problems (CSP). A search problem
is defined as the problem of finding a satisfying assignment for a set (formula) of constraints acting on $n$ 
(qu)bits. In general, each constraint acts on a number of bits less than or equal to $n$. The search 
will terminate whenever the ``program'' finds an assignment, or 
assignments, that satisfy simultaneously all the constraints of the CSP formula. For example, the 
oracle search over an unstructured database corresponds to the problem of finding an assignment that 
satisfies all the constraints of the predefined formula, whose constraints are all $n$ bit constraints. 
Even though the number of qubits is fixed, by varying the length or the type of constraints the CSP can 
cover different types of problems. 

For a structured database (problem) we can divide the initial search problem of $n$ qubits into 
subproblems with smaller number of variables. One can choose to test first if an assignment to 
$n_A$ variables out of the $n$ variables satisfies all the constraints acting uniquely on this 
subset of variables, labeled $A$. The result of this test is  kept in a register. A similar test is then 
applied to the remaining bits, that we label as subset $B$. And finally one must  
nest both tests to find a global solution for the initial search problem. This is 
what is meant by a structured or nested search.

More specifically, assume that the initial search problem admits a number $M_{AB}$ of solutions. In a quantum search 
problem, a solution or a satisfying assignment is a state in the Hilbert space of the $n$ qubits. 
The set of solutions we thereafter denote by  
\begin{equation}
{\cal M}_{AB}=\left\{ |m_i\rangle /i=1,2,\cdots, M_{AB}\right\}\; .
\end{equation} 
All elements of this set satisfy the predefined formula of constraints denoted generically 
by $\left\{C_{AB}\right \}$. For our patching strategy, the constraints $\left\{C_{AB}\right\}$ are 
classified 
as
\begin{itemize}
\item $\left\{C_A\right\}$: the set of constraints acting only on the subset $A$. In 
general, there exist different solutions satisfying all the constrains of this set. The 
set of possible solutions is  
\begin{equation}
{\cal M}_{A}=\left\{ |m_i^A\rangle /i=1,2,\cdots,  M_{A}\right\}\; .
\end{equation} 
\item $\left\{C_B\right\}$: the set of constraints acting only on the subset $B$. Again, 
different solutions can be found to satisfy all the constraints of this set. The 
set of possible solutions is  
\begin{equation}
{\cal M}_{B}=\left\{ |m_i^B\rangle /i=1,2,\cdots, M_{B}\right\}\; .
\end{equation} 
\item $\left\{C_{A/B}\right\}=\left\{C_{AB}\right \}-\left\{C_{A}\right \}-\left\{C_{B}\right \}$ is the set of constraints acting simultaneously on $A$ and $B$\footnote{This procedure is analogous to the {\it resolution} method used in Davis-Putnam 
procedure \cite{DavisP1}. Hence $\left\{C_{A/B}\right\}$ can then be called the {\it reduction} of  $\left\{C_{AB}\right \}$ by $A$ and $B$.}.  
\end{itemize}
Some elements of ${\cal M}_{A}$ may not satisfy one or more constraints in 
$\left\{C_{A/B}\right \}$, those elements can not give global solutions, so they are 
{\it Not Solutions}. The set of such elements is denoted by ${\cal M}_A^{NS}$. On the other hand, 
the set of elements of ${\cal M}_{A}$ satisfying all the constraints in $\left\{C_{A/B}\right \}$, 
and hence giving global solutions, will be denoted by ${\cal M}_{A}^S$. Therefore  
\begin{equation}
{\cal M}_A={\cal M}_A^S\cup {\cal M}_A^{NS}\; . 
\end{equation} 
Similarly we have 
\begin{equation}
{\cal M}_B={\cal M}_B^S\cup {\cal M}_B^{NS}\; . 
\end{equation} 
So, an element $|m_i\rangle\in {\cal M}_{AB}$ can be written as 
\begin{equation}
|m_i\rangle=|m_A^S\rangle\otimes |m_B^S\rangle\; ,
\end{equation} 
where $|m_A^S\rangle$ and $|m_B^S\rangle$ are elements of ${\cal M}_A^S$ and 
${\cal M}_B^S$ respectively. 

It is evident that for a given state $|m_A^S\rangle$ there could be different states 
$|m_B^S\rangle$ such that their product is an element of ${\cal M}_{AB}$. At some stages, the authors 
of \cite{CerfGW1} and \cite{RolanC3} made the assumption that for a given 
$|m_A^S\rangle$ there is only one $|m_B^S\rangle$ such that their product is a global 
solution. Although one may argue that such an assumption is justified for hard search problems, where 
there is just one global solution, it is unlikely that it holds for a generic search problem.  
This assumption is not needed for our algorithm, and hence our algorithm applies to a 
broader class of search problems.

%%%%%%%%%%%%%%%%%%%%%%%%%%%%%%%%%%%%%%%%%%%%%%%%%%%%%%%%%%%%%%%%%%%%%%%%%%%%%%%%%%%%%%
\section{Nested algorithm}
%%%%%%%%%%%%%%%%%%%%%%%%%%%%%%%%%%%%%%%%%%%%%%%%%%%%%%%%%%%%%%%%%%%%%%%%%%%%%%%%%%%%%%
The present algorithm consists of two stages. In the first stage (stage $I$) we evolve 
adiabatically the quantum system from a state which is the ground state of a 
Hamiltonian that can be easily obtained to a state which is a product  
of states  of ${\cal M}_A$ and states of ${\cal M}_B$. As previously mentioned, this product is not 
yet a global solution. In stage 
$II$ a global search, similar to what is labeled as stage C in \cite{RolanC3}, is applied 
to the output of stage $I$ to rotate it towards an element of ${\cal M}_{AB}$. 
The output of stage $II$ is a global solution of the problem.  

%%%%%%%%%%%%%%%%%%%%%%%%%%%%%%%%%%%%%%%%%%%%%%%%%%%%%%%%%%%%%%%%%%%%%%%%%%%%%%%%%%%%%%
\subsection{Stage $I$}
%%%%%%%%%%%%%%%%%%%%%%%%%%%%%%%%%%%%%%%%%%%%%%%%%%%%%%%%%%%%%%%%%%%%%%%%%%%%%%%%%%%%%%
In this stage we use the procedure defined in \cite{AhrenKKZD2}. The Hamiltonian 
is split into two parts, each acting on one of the Hilbert spaces ${\cal H}_A$, of 
dimension $N_A=2^{n_A}$, and ${\cal H}_B$, of dimension $N_B=2^{n_B}$, spanned by  
subsets $A$ and $B$ respectively.\\ 
The most convenient initial state is the equal superposition of all possible pure 
states of the system  
\begin{eqnarray}    
|\Psi(t=0)\rangle 
%|\Psi_0\rangle=\frac{1}{\sqrt{N}}\sum\limits_{i\in {\cal H}_A\otimes{\cal H}_B}|i\rangle
&=&\frac{1}{\sqrt{N_A}}\sum\limits_{i_A\in {\cal H}_A}|i_A\rangle\otimes\frac{1}{\sqrt{N_B}}\sum\limits_{i_B\in {\cal H}_B}|i_B\rangle\;\nonumber\\ 
&\equiv & |\Psi_{0A}\rangle\otimes|\Psi_{0B}\rangle\; ,
\end{eqnarray}
where $N=2^n=N_AN_B$. $|\Psi_0\rangle$ is the ground state of the Hamiltonian
\begin{equation} 
H_0=(I_A-|\Psi_{0A}\rangle\langle\Psi_{0A}|)\otimes I_B+ I_A\otimes(I_B-|\Psi_{0B}\rangle\langle\Psi_{0B}|)\; .
\end{equation}
The initial Hamiltonian should evolve adiabatically into a final Hamiltonian which has the 
ground state 
\begin{equation}
\frac{1}{\sqrt{M_A}}\sum\limits_{m_A\in {\cal M}_A}|m_A\rangle\otimes\frac{1}{\sqrt{M_B}}\sum\limits_{m_B\in {\cal M}_B}|m_B\rangle\equiv |\Psi_{fA}\rangle\otimes|\Psi_{fB}\rangle \; .
\end{equation}
A possible final Hamiltonian is   
\begin{equation}
H_f^I=(I_A-|\Psi_{fA}\rangle\langle\Psi_{fA}|)\otimes I_B+ I_A\otimes(I_B-|\Psi_{fB}\rangle\langle\Psi_{fB}|)\; .
\end{equation}
We have chosen $H_0$ and $H_f^I$ so that the evolution is decoupled in ${\cal H}_A$ and ${\cal H}_B$. 
The time dependent Hamiltonian evolving $H_0$ to $H_f^I$ in a time $T_I$ 
is the linear combination 
\begin{equation}
 H(t)=(1-s(t))\,H_0 + s(t)\,H_f^I\; \equiv f(s)\;H_0+ g(s)\, H_f^I, 
\label{adiab_ham}
\end{equation}
where $s(t)$ is chosen such that $s(0)=0$ and $s(T_I)=1$. When the adiabaticity condition  
holds, the 
initial ground state 
$|\Psi_0\rangle$ will evolve slowly in time $T_I$ to the ground state of the final 
Hamiltonian. The meaning of ``slow'' evolution is quantified by the adiabatic theorem \cite{Bransden}, which states that the accuracy with which the system remains in its 
instantaneous ground state is equal to the sum of the ratio of the transition matrix element from the ground 
state, with energy $E_0$, to any other state, with energy $E_i$, over the fourth power of the 
``radiated'' energy $\omega_{0i}=E_0-E_i$, {\it i.e.} 
\begin{equation}\label{DADT}
 \sum_{i} \frac{\left |\langle E_i |\frac{dH}{dt}|E_0\rangle  \right |^2}{\omega_{0i}^4}
 \leq \epsilon^2\, . 
\end{equation}
This gives a lower bound on the evolution time $T_I$. 

The eigenvalues and eigenfunctions  of the 
Hamiltonian $H(t)$ can be calculated analytically. We find that the only nonvanishing 
transition probabilities from the ground state are to the two lowest energy levels: 
\begin{itemize}
\item $E_A=E_1^A+E_0^B$, {\it i.e.} the state where the subsystem $A$ is in its first excited state, and the 
subsystem $B$ is in its ground state 
\item  $E_B=E_0^A+E_1^B$, {\it i.e.} the state where the subsystem $A$ is in its ground 
state, and $B$ is in its first excited state. 
\end{itemize}
The corresponding transition matrix elements are given by   
\begin{equation}
\left |\langle E_i |\frac{dH}{dt}|E_0\rangle  \right |^2 =\left(\frac{ds}{dt}\right)^2 \frac{M_i^2/N_i^2}{\omega_i^2}\left(\frac{N_i}{M_i}-1\right)\equiv \frac{\xi_i^2}{\omega_i^2}\left(\frac{ds}{dt}\right)^2\; ,\; i=A,B \; ,
\label{eq:transition}
\end{equation} 
where 
\begin{equation}
 \omega_i=\sqrt{(f-g)^2+\frac{4M_i}{N_i}fg}=\sqrt{(1-2s)^2+4\frac{M_i}{N_i}s(1-s)}\, 
\end{equation}
is the gap (energy difference) between the first excited state and the ground state of the 
$i$-th subsystem. \\
By integrating eq.~(\ref{DADT}) over $s$ using eq.~(\ref{eq:transition}) the lower bound, or the 
minimal time, will be    
\begin{equation}\label{time}
 T_I= \frac{1}{\epsilon}\int_0^1 \, ds\,\sqrt{\sum_{i=A,B} \frac{\xi_i^2}{\omega_i^6}}\; .
\label{eq:time}
\end{equation}

%%%%%%%%%%%%%%%%%%%%%%%%%%%%%%%%%%%%%%%%%%%%%%%%%%%%%%%%%%%%%%%%%%%%%%%%%%%%%%%%%%%%%%
%\subsubsection{Stage $I$ and Cerf {\it et al.} algorithm} 
%%%%%%%%%%%%%%%%%%%%%%%%%%%%%%%%%%%%%%%%%%%%%%%%%%%%%%%%%%%%%%%%%%%%%%%%%%%%%%%%%%%%%%
Note that the {\it parallel} evolution, in the two Hilbert spaces ${\cal H}_A$ and ${\cal H}_B$, of stage $I$ 
is the major feature that distinguishes our algorithm from that of \cite{CerfGW1}. Stage $I$ replaces 
the two-stages evolution ({\it sequential} evolution) used in \cite{CerfGW1}. A priori the 
sequential evolution seems more efficient, as one can use the result of the first stage to 
eliminate some of the no-good assignments in the second stage. However the algorithm used in 
\cite{CerfGW1} is not a {\it tree-like} search procedure, 
{\it i.e.} it does not pick an order in which it instantiates the variables. Hence the 
algorithm of \cite{CerfGW1} does not eliminate ``bad'' trees. Therefore replacing the 
sequential by the parallel evolution will not affect dramatically the time cost of the algorithm. 
On the other hand, as mentioned before , the symmetric parallel evolution allows one to analyse a 
broader set of structured search algorithms.

%%%%%%%%%%%%%%%%%%%%%%%%%%%%%%%%%%%%%%%%%%%%%%%%%%%%%%%%%%%%%%%%%%%%%%%%%%%%%%%%%%%%%  
\subsection{Stage $II$} 
%%%%%%%%%%%%%%%%%%%%%%%%%%%%%%%%%%%%%%%%%%%%%%%%%%%%%%%%%%%%%%%%%%%%%%%%%%%%%%%%%%%% 
The output state of stage $I$ can be written as the sum of a state encoding the solution to the 
search problem and a residual part, which is not a solution: 
\begin{equation}
|\Psi_{fA}\rangle\otimes|\Psi_{fB}\rangle=\sqrt{\frac{M_AM_B-M_{AB}}{M_AM_B}}|\Psi^{NS}\rangle+\sqrt{\frac{M_{AB}}{M_AM_B}}|\Psi^S\rangle\; ,
\end{equation}
where the ``solution state'' is 
\begin{equation}
|\Psi^S\rangle=\frac{1}{\sqrt{M_A^SM_B^S}}\sum\limits_{m_A^S\in {\cal M}_A^S}|m_A^S\rangle\otimes\sum\limits_{m_B^S\in {\cal M}_B^S}|m_B^S\rangle=\frac{1}{\sqrt{M_{AB}}}\sum\limits_{m_i\in {\cal M}_{AB}}|m_i\rangle
\end{equation}
and we have used the fact that $M_A^SM_B^S=M_{AB}$.

At this stage one is inclined to start again an adiabatic evolution from a Hamiltonian whose ground 
state is $|\Psi_{fA}\rangle\otimes|\Psi_{fB}\rangle$, {\it e.g.} 
\begin{equation} 
H_i=1-|\Psi_{fA}\rangle\langle\Psi_{fA}|\otimes|\Psi_{fB}\rangle\langle\Psi_{fB}|
\end{equation}
to a final Hamiltonian whose ground state is $|\Psi^S\rangle$, {\it e.g.} 
\begin{equation}
H_f=1-|\Psi^S\rangle\langle\Psi^S|\; , 
\end{equation}
in time $O(\sqrt{M_AM_B/M_{AB}}\;)$. 
Unfortunately, this is not possible since the initial Hamiltonian $H_i$ is not accessible. Following \cite{RolanC3} a 
global search can be applied to nest between the 
two subsets $A$ and $B$. The global search is achieved through the following procedure
\begin{itemize}
\item Stage $I$ is approximated by an evolution operator $U$ such that 
$|\Psi_{fA}\rangle\otimes|\Psi_{fB}\rangle \approx U |\Psi_{0A}\rangle\otimes|\Psi_{0B}\rangle$. 
\item Hence $H_i\approx UH_0U^{\dagger}$, note that $H_i$ is ``replaced'' by $H_0$ which 
is easily accessible.  
\item The adiabatic evolution from $H_i$ to $H_f$ is implemented on quantum circuits. 
The continuous evolution is replaced by a ``step evolution'' over intervals of time. In 
each step, at a given time $t$, we use the approximation 
\begin{equation}
e^{-iH_it}\approx U e^{-iH_0t}U^{\dagger} 
\end{equation}
{\it i.e.} a backword evolution in time $T_I$, an application 
of $\exp(-iH_0t)$, then a forward evolution in time $T_I$. This step requires a time $O(T_I)$.
\end{itemize}
The number of steps is chosen to minimize the  error involved in the discretization needed during the 
global evolution. The number of iteration is at least 
(see \cite{RolanC3}) of the order of $\sqrt{M_AM_B/M_{AB}}$. So that the total running time   
needed to get a global solution, is proportional to 
\begin{equation}
T=T_I\sqrt{\frac{M_AM_B}{M_{AB}}}\; =T(M_A,M_B,M_{AB},n,n_A/n).
\end{equation}
This expression is not yet the final answer, since there is still the problem of determining 
the values of $M_{A,B,AB}$. This will be discussed in the next section.

%%%%%%%%%%%%%%%%%%%%%%%%%%%%%%%%%%%%%%%%%%%%%%%%%%%%%%%%%%%%%%%%%%%%%%%%%%%%%%%%%%%%%  
\section{Complexity analysis} 
%%%%%%%%%%%%%%%%%%%%%%%%%%%%%%%%%%%%%%%%%%%%%%%%%%%%%%%%%%%%%%%%%%%%%%%%%%%%%%%%%%%% 
The total running time $T$ is a function of $M_A,M_B,M_{AB},n$ and $x=n_A/n$, among which $n$ is 
the only input parameter, and $x$ will be chosen to minimize $T$. Hence 
we are left with the three parameters $M_A,M_B$ and $M_{AB}$. A priori, these parameters are 
problem dependent and are difficult to determine. However, since we are interested in getting 
general results which depend as little as possible on the details of the problem, we 
can approximate these quantities using a complexity analysis. 

Computational 
complexity theory studies the quantitative laws which govern computing. It seeks a comprehensive 
classification of problems by their intrinsic difficulty and an understanding of what makes these 
problems hard to compute. The simplified average complexity analysis derived in 
\cite{WilliH1,CerfGW1} can be adapted for our generic algorithm. The major simplification made in 
\cite{CerfGW1} is the approximation of independent no-good assignments\footnote{For most CSP 
there are classes of no-good assignments that can be deduced from each other. For example, 
one can use proposional reasoning to generate some no-goods from others, see for example 
chapter two of \cite{Bacchus}.}. This allows for a relatively 
easy combinatoric analysis. Following \cite{CerfGW1} the unknown parameters are estimated by 
\begin{equation}
M_j\approx 2^{n_j-n\alpha(n_j/n)^k}\quad {\rm for}\quad j=A,B, AB\quad {\rm with}\;\;\; n_{AB}=n\; ,    
\end{equation}
where $k$ is the number of variables per constraint, assuming a constant length constraint. 
$\alpha$ represents the average difficulty of the problem, it characterizes 
the average number of no-good ground instances per variable. Hardest problems are found near a 
critical\footnote{This critical behavior is similar to a phase transition in condensed matter 
physics. Similar phase transition features were observed for random K-SAT problems 
\cite{MitchSL1,KirkS1}.} value $\alpha=\alpha_c=1$. The critical $\alpha$ is obtained from 
the number of solutions  
$$M_{AB}\approx N^{1-\alpha}$$ 
for $\alpha < 1$, $M_{AB}$ is large, {\it i.e.} the problem is under-constrained  and easy 
to solve, while for $\alpha > 1$, $M_{AB}<1$, {\it i.e.} the problem is over-constrained and 
probably has no solution. Finally for $\alpha=1$ we have one solution, this is the definition 
of the critical value. 

Therefore, the optimization procedure reduces to a  minimization of the running time as a function of 
the ratio $x$ for different values of $\alpha$ and $k$.

%%%%%%%%%%%%%%%%%%%%%%%%%%%%%%%%%%%%%%%%%%%%%%%%%%%%%%%%%%%%%%%%%%%%%%%%%%%%%%%%%%%%%  
\subsection{Approximate and exact scalings} 
\begin{figure}
\centerline{\resizebox*{!}{5cm}{\includegraphics{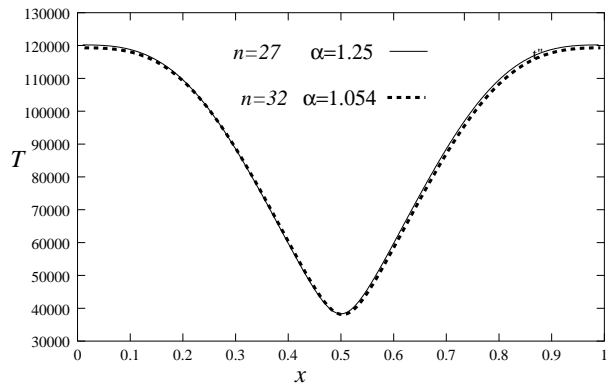}}}
\caption{\label{fig:var-alpha-n} The running time $T$ as a function of $x$, for 
$(k=3,\alpha=1.054,n=32)$, $(k=3,\alpha=1.25,n=27)$, such that $n\alpha$ is approximately constant. 
This supports the effective parameterization in terms of $n\alpha$.}
\end{figure}

\begin{figure}
\centerline{\resizebox*{!}{5cm}{\includegraphics{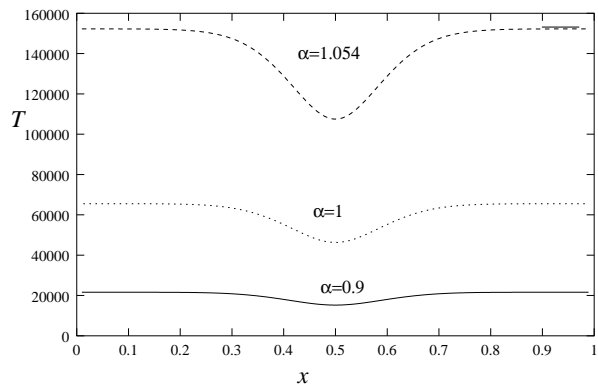}}}
\caption{\label{fig:var-alpha} The running time $T$ as a function of $x$, for 
$(k=5,\alpha=0.9,1,1.054, n=32)$. The slightest variation of $\alpha$ affects
dramatically the running time, which
reflects the critical behavior predicted by the complexity 
analysis.}
\end{figure}

\begin{figure}[h]
\centerline{\resizebox*{!}{5cm}{\includegraphics{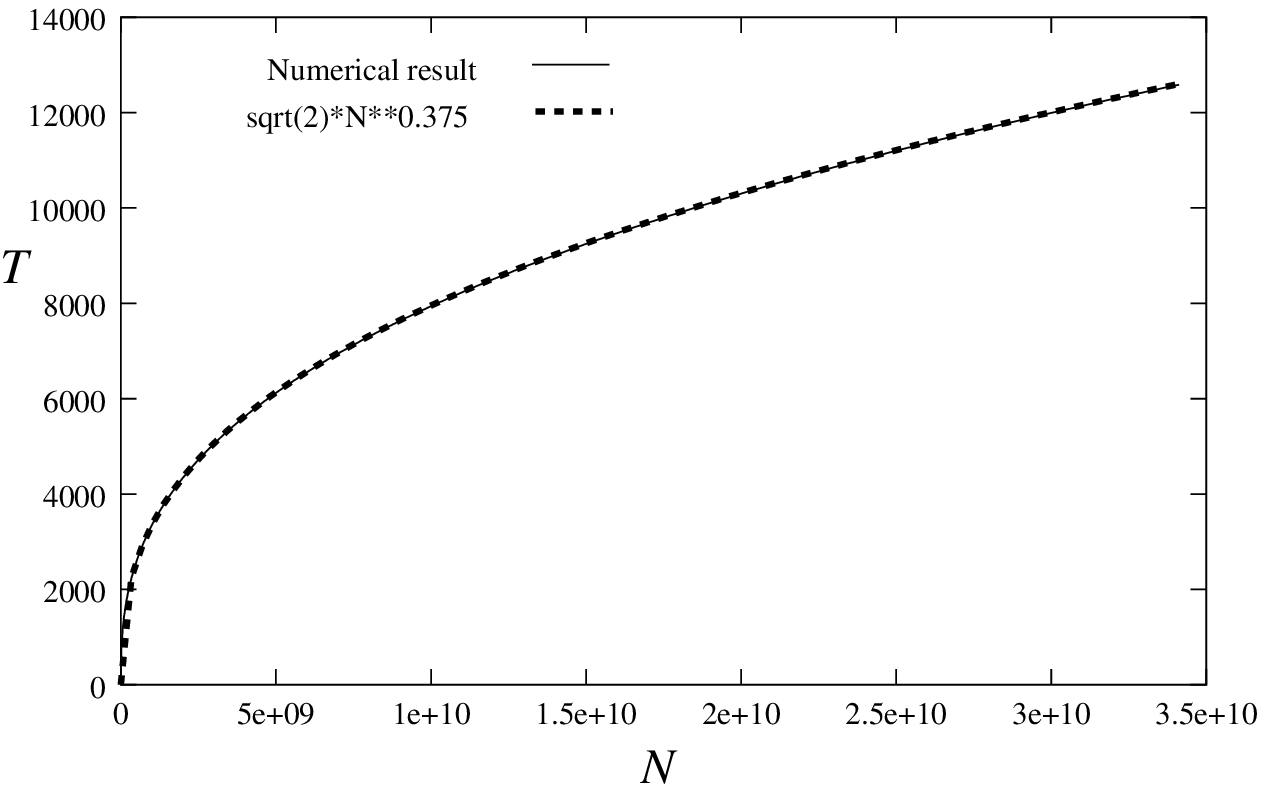}}}
\caption{\label{fig:scaling} The running time $T$ as a function of $N$, for 
$(k=2,x=0.5)$. The numerical result matches the approximate analytical scaling: $T\sim O(N^{0.375})$.}
\end{figure}

%%%%%%%%%%%%%%%%%%%%%%%%%%%%%%%%%%%%%%%%%%%%%%%%%%%%%%%%%%%%%%%%%%%%%%%%%%%%%%%%%%%% 
The running time $T$ can not be calculated analytically. However 
by the following simple arguments it is possible to give an estimate for $T$:   
\begin{itemize}
\item We first recognize
that our algorithm is inefficient if $M_i/N_i\sim 1$. This can be easily 
seen by remembering that $N_i$ is the number of possible assignments in space $A$ or $B$ and $M_i$ 
is the number of solution in that space. For $M_i\sim N_i$ any assignment will become a solution, then, 
one step (picking up any state) is sufficient to find a solution in the subspace $i$. This is reflected 
through the limit 
$$\xi_i=\frac{M_i}{N_i}\sqrt{\frac{N_i}{M_i}-1}\rightarrow 0 \quad {\rm when}\;\; M_i\rightarrow N_i\; .$$  
Therefore, the first stage does not eliminate any no-good assignment and consequently the second stage  
will take $\sqrt{N}$ steps to  give a solution, {\it i.e.} as if the search is an oracle search. But what does 
this means? \\
Using the average complexity analysis of the previous section we write:
$$ M_A/N_A=2^{-n\alpha x^k}\; .$$ 
Taking $M_A/N_A\sim 1$ implies that $ n\alpha x^k\sim 1$ or $k\sim \ln(n\alpha)/\ln(1/x)$, on 
the other hand we 
expect\footnote{If $x=0$ or 1, the first stage is again useless as there is no division of the 
initial space.} that optimization will give a value of $x$  not far from $1/2$, hence for large 
values of $n$ and near the critical $\alpha=1$ our algorithm breaks down for $k\sim \ln(n)$. But 
the larger the $k$ is, the closer we are to the unstructured search where each constraint is an $n$ 
qubit constraint. 
\item From the previous remark we deduce that our algorithm is effective as long as $M_i/N_i\ll 1$. 
\item The running time of stage $I$, given in eq.~(\ref{eq:time}), exhibits a near singular behavior 
for  $M_i/N_i\ll 1$. This near singularity occurs when the integration variable $s$ is close to the value for 
which $f(s)=g(s)$, 
{\it i.e. $s=1/2$}.    
\end{itemize}
\begin{figure}
\centerline{\resizebox*{!}{5cm}{\includegraphics{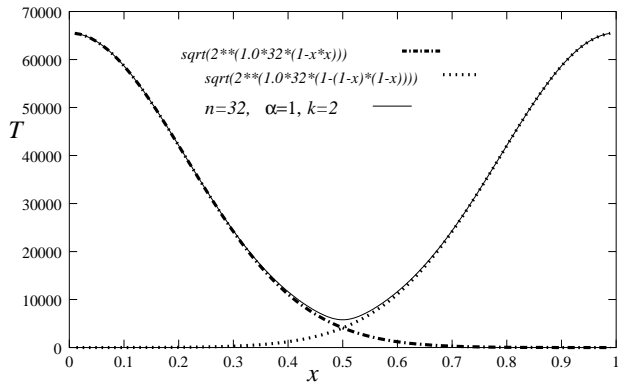}}}
\caption{\label{fig:num-anal} The numerical value of the running time $T$ as a function of $x$, for 
$(k=2,\alpha=\alpha_c=1,n=32)$, compared to the analytical estimate of $T$. Similar results 
are obtained for different values of $k$.}
\end{figure}
These remarks imply that, for a given $x$, $T_I$ can be approximated using the near singular 
behavior to obtain
\begin{equation}
T_I(x)\sim \sqrt{{\rm Max} \left( \frac{N_A}{M_A}, \frac{N_B}{M_B}\right)}\; .
\end{equation}
In other words, the time $T_I$ scales with the square root of the largest value of the dimension 
of each Hilbert space divided by the number of possible solutions in that space.   
Multiplying by the number of iterations of stage $II$, the total running time reads  
\begin{equation}
T(x)\sim \sqrt{{\rm Max} \left( \frac{N_AM_B}{M_{AB}}, \frac{N_BM_A}{M_{AB}}\right)}\; .
\end{equation}
%For $\alpha\approx 1$, $M_{AB}\approx 1$ and  
Using the average complexity analysis we get 
\begin{equation}
 T(x)\sim \sqrt{{\rm Max} \left( N^{\alpha-\alpha(1-x)^k}, N^{\alpha-\alpha x^k}\right)}\; .
\end{equation}
In figure~\ref{fig:num-anal} we compare this approximate formula with the time calculated numerically. 
The two results match extremely well for all values of $x$. \\
\begin{figure}
\centerline{\resizebox*{!}{5cm}{\includegraphics{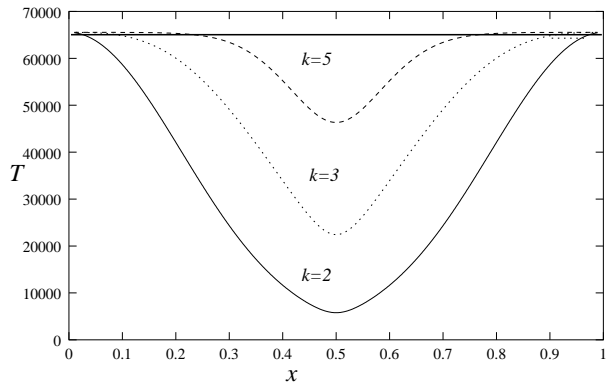}}}
\caption{\label{fig:var-k} The running time $T$ as a function of $x$, for 
$(k=2, 3,5,\alpha=\alpha_c=1,n=32)$. The optimal time is attained at $x=1/2$. The horizontal 
line is the $\sqrt{N}$ scaling. It is evident that the larger $k$ is the closer we are 
to the horizontal line, {\it i.e.} to the oracle search.}
\end{figure}
The time $T$ is then optimal for $N^{\alpha-\alpha(1-x)^k}= N^{\alpha-\alpha x^k}$, {\it i.e.}
$x=1/2$, which is compatible with the symmetric nature of the algorithm. This result agrees 
with the numerical results presented in figure~\ref{fig:var-k}, where $x=1/2$ is found to 
give the optimal time for different values of $k$. The plot 
also illustrates that the larger $k$ is the closer we are 
to an oracle type search, which is what we 
predicted before.
Hence the optimal time is given by 
\begin{equation}
T_{min}=T(x=1/2)\sim N^{\alpha/2-\alpha/2^{k+1}}\sim 2^{n\alpha(1/2-1/2^{k+1})}\; . 
\label{eq:scaling}
\end{equation}
We conclude that the parameter of interest in these
considerations is $n\alpha$. In figure~\ref{fig:var-alpha-n} 
we plot the 
running time as a function of $x$ for different values of $\alpha$ and $n$ keeping their product 
constant. The fact that the different plots coincide
strongly supports the scaling obtained from the 
analytical estimate. \\ 
Another important aspect is the (sub)-exponential growth of $T$ as a function of $\alpha$. This can 
be seen either from eq.~(\ref{eq:scaling}) or from the ``critical'' behavior depicted in 
figure~\ref{fig:var-alpha}. \\   
Finally we consider the special case $(k=2, \alpha=1)$.
In this case the running time is $O(N^{3/8})$, which   
is better than the classical running time $O(N^{0.5})$ but less efficient than the result obtained 
using the algorithm of Cerf {\it et al.} which gives $O(N^{0.31})$. The $O(N^{0.375})$ scaling is 
compared with the numerical result in figure~\ref{fig:scaling}.   

%%%%%%%%%%%%%%%%%%%%%%%%%%%%%%%%%%%%%%%%%%%%%%%%%%%%%%%%%%%%%%%%%%%%%%%%%%%%%%%%%%%%%  
\section{Conclusions} 
%%%%%%%%%%%%%%%%%%%%%%%%%%%%%%%%%%%%%%%%%%%%%%%%%%%%%%%%%%%%%%%%%%%%%%%%%%%%%%%%%%%% 
We have presented a new adiabatic quantum algorithm for searches over structured databases. 
Our algorithm is constructed with minimum assumptions about the nature of the database and the 
specificity of the search problem. Moreover, it results in a significant potential speedup 
over its classical counterpart. 

The new algorithm is in fact ``quasi''-adiabatic, it requires the use of quantum circuits 
at some stages. An interesting topic for future work would be to construct a ``pure'' adiabatic 
quantum algorithm. Another important issue is to explore the possibility of using smart classical 
strategies such as {\it back tracking} and {\it constraint propagation}, and to see the effect of such strategies on quantum 
interference and coherence. These and other related issues will be addressed elsewhere.

\appendix

%%%%%%%%%%%%%%%%%%%%%%%%%%%%%%%%%%%%%%%%%%%%%%%%%%%%%%%%%%%%%%%%%%%%%%%%%%%%%%%%%%%%%  
\section{Multi-nesting verses multi-partition} 
%%%%%%%%%%%%%%%%%%%%%%%%%%%%%%%%%%%%%%%%%%%%%%%%%%%%%%%%%%%%%%%%%%%%%%%%%%%%%%%%%%%% 
The nesting procedure can be applied to subsets $A$ and $B$, resulting  in higher level 
nesting 
or multi-nesting \cite{CerfGW1}. In multi-nesting stages $I$ and $II$ are applied to 
subsets $A$ and $B$ separately. Then stage $II$ is applied to nest $A$ and $B$. Then the 
time $T_I$ is replaced by a shorter time. More nesting can be used to enhance the 
effectiveness of the structured search. 

Another alternative is to modify stage $I$ by splitting the $n$ variables into more than 
two subset then 
\begin{equation}
 T_I= \frac{1}{\epsilon}\int_0^1 \, ds\,\sqrt{\sum_{i=A,B,C,D,...} \frac{\xi_i^2}{\omega_i^6}}
\end{equation}
and hence applying the procedure of stage $II$ we get 
\begin{equation}
T=T_I\sqrt{\frac{\prod\limits_{i=A,B,C,D,...}M_i}{M_{AB}}}\; =T(M_A,M_B,M_C,...,M_{AB},n,n_A/n).
\end{equation}
An average complexity analysis can then be applied to obtain an estimate for $T$.

\end{document}